\documentclass[aps,pra,reprint,superscriptaddress]{revtex4-1}

\usepackage{graphicx}
\usepackage{amsmath}
\usepackage{amsfonts}
\usepackage{multirow}
\usepackage{bbold}
\usepackage{makecell}
\usepackage{tabu}

\begin{document}

\title{Anyonic quasiparticles of hardcore anyons}

\author{Julia Wildeboer}
\affiliation{Max-Planck-Institut f{\"u}r Physik komplexer Systeme, D-01187 Dresden, Germany}
\affiliation{Department of Physics, Arizona State University, Tempe,
Arizona 85287-1504, USA}
\affiliation{Department of Physics and Astronomy, University of Kentucky, 505 Rose Street, Lexington, KY 40506}

\author{Aniket Patra}
\affiliation{Max-Planck-Institut f{\"u}r Physik komplexer Systeme, D-01187 Dresden, Germany}

\author{Sourav Manna}
\affiliation{Max-Planck-Institut f{\"u}r Physik komplexer Systeme, D-01187 Dresden, Germany}

\author{Anne E. B. Nielsen}
\altaffiliation{On leave from Department of Physics and Astronomy, Aarhus University, DK-8000 Aarhus C, Denmark.}
\affiliation{Max-Planck-Institut f{\"u}r Physik komplexer Systeme, D-01187 Dresden, Germany}

\begin{abstract}
Strongly interacting topologically ordered many-body systems consisting of fermions or bosons can host exotic quasiparticles with anyonic statistics. This raises the question whether many-body systems of anyons can also form anyonic quasiparticles. Here, we show that one can, indeed, construct many-anyon wavefunctions with anyonic quasiparticles.  The braiding statistics of the emergent anyons are different from those of the original anyons. We investigate hole type and particle type anyonic quasiparticles in Abelian systems on a two-dimensional lattice and compute the density profiles and braiding properties of the emergent anyons by employing Monte Carlo simulations.
\end{abstract}

\maketitle

\begin{figure*}
\includegraphics[trim={0.7cm 6cm 0.7cm 3cm},clip,scale=0.48]{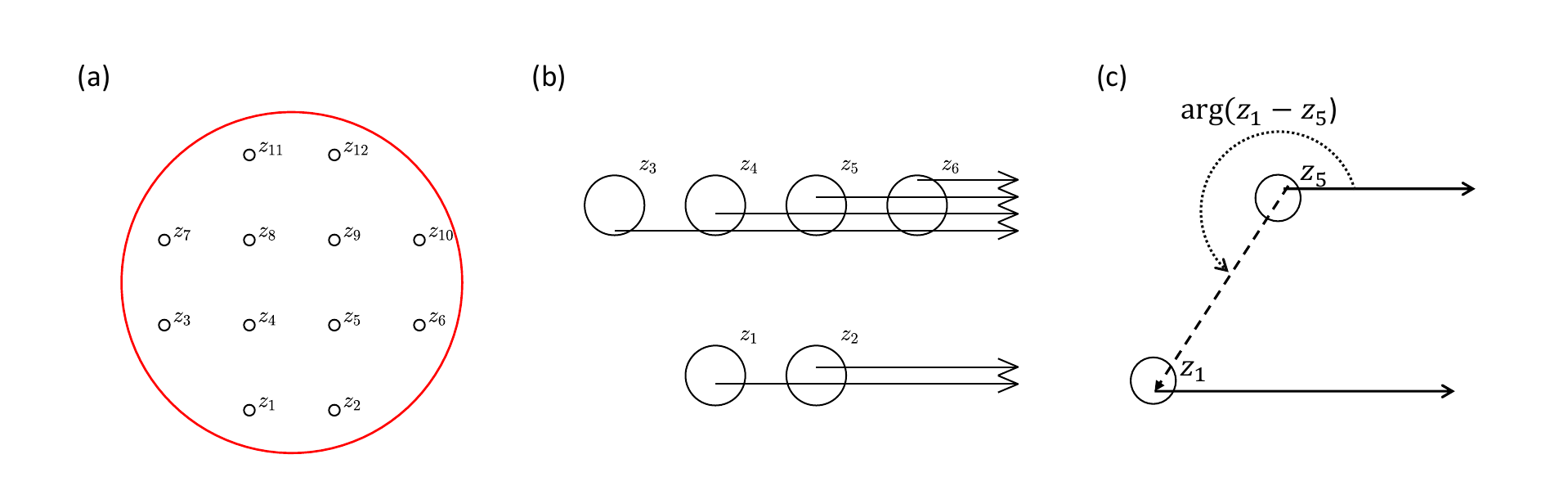}
\caption{Choice of a definite branch for the anyon wavefunction \eqref{state}. We consider a square lattice of lattice spacing $\sqrt{2\pi}$ with a roughly circular edge obtained by cutting the lattice along the red circle of radius $R=0.01 + 2\sqrt{2\pi}$. We always number the lattice sites as shown in (a), following the rule that $j>k$ if $\mathrm{Im}(z_j)>\mathrm{Im}(z_k)$, and if $\mathrm{Im}(z_j)=\mathrm{Im}(z_k)$, then $j>k$ if $\mathrm{Re}(z_j)>\mathrm{Re}(z_k)$. The individual branch cuts (solid straight arrows) corresponding to each lattice coordinate $z_{i}$ (we only show a few of them) are shown in (b). For sites on the same line, the branch cut is displaced higher up, the further the point is to the right. We consider a typical $z_{i} - z_{j}$ (dashed arrow) in (c) and show how to uniquely choose the $\textrm{arg}(z_{i} - z_{j})$ that lies between $0$ and $2\pi$.}
\label{LatticeOrdering}
\end{figure*}

\section{Introduction}

Quantum statistics is an important concept for gaining insight into numerous observed collective phenomena in nature. It was first realized more than forty years ago, in a seminal paper by Leinaas and Myrheim \cite{LM}, that upon restricting a system to two spatial dimensions, there appears another type of identical particles in addition to the usual bosons and fermions. These particles, which satisfy braid statistics instead of permutation statistics, were coined as \textit{anyons} by Wilczek \cite{FW1, FW2}. Anyons come in two flavors: Abelian and non-Abelian. The former have the property that under continuous adiabatic exchange of two anyons, the many-body wavefunction acquires a complex phase factor $e^{i\phi} \neq \pm 1$, while the latter exhibit even more striking features that make those potential candidates for quantum computations \cite{RMP}.

The literature presents a body of works to approach the physics of anyons from various angles.  An important example of an intensively studied system that naturally hosts anyons as quasiparticle excitations is the fractional quantum Hall system \cite{tsui,lau1983}. Laughlin explained the fractional quantum Hall system at filling fraction $1/3$ as an incompressible quantum liquid with fractionalized low-energy excitations \cite{lau1983}. More recently, anyons have been investigated in lattice fractional quantum Hall systems \cite{aebnbraid, nees}, and there are several proposals for realizing these systems in ultracold atoms in optical lattices \cite{Others49}.

The contemporary literature provides examples of two-dimensional strongly correlated systems consisting of many interacting bosons or fermions that have anyonic quasiparticle excitations \cite{RMP}, and much work has been done to investigate and classify the different types of anyons that appear in these systems \cite{wen1990, classification, smatrix}. Instead of studying systems with isolated anyonic excitations, one can also study systems with many anyons. Such ideas have, e.g., been used in Haldane's hierarchy construction \cite{HierHal} to propose trial states for fractional quantum Hall systems at Landau level filling fractions other than $1/3$ -- e.g.\ $2/5$ and $2/7$.

Another possibility is to construct systems, in which the constituent particles themselves behave like anyons. This can, e.g., be achieved by considering individual hardcore anyons as hardcore bosons or fermions with attached point flux tubes \cite{FWgauge}. One can then construct a continuum Hamiltonian for the hardcore boson (fermion) part of the many-anyon system \cite{lund1}. The point flux tubes are included in the ``statistical" part of the magnetic vector potential in the many-body Hamiltonian with the help of a singular gauge transformation. One can obtain the exact ground state basis for such a continuum many-anyon Hamiltonian, which is entirely confined to the lowest Landau level \cite{polychron1, ouvry1, ouvry2}. A recent work has introduced variational ans\"{a}tze for the ground state of the above Hamiltonian \cite{lund2, lund3}. One such ansatz was shown to be the same as the Read-Rezayi state \cite{RR}. The possible existence of non-Abelian emergent anyonic excitations over such ground state trial wavefunctions was then indicated using the special clustering properties of certain symmetric polynomials.

A lot of work has been done on the statistical mechanics of the many-anyon system \cite{kharebook}. A fractional exclusion principle (generalized version of the Pauli exclusion principle) for such problems allows one to obtain an equation of state \cite{frachal, fracwu, mitra, nayak}. A representation of the Hilbert space for the multi-anyon state was constructed using the braided tensor categories, which then provides a different perspective for the fractional exclusion statistics \cite{goldin}. Instead of resorting to the fractional exclusion principle, starting from the continuum many-anyon Hamiltonian, perturbative results elucidate the complexity of the equation of state \cite{DS, ouvry4}. One can even derive an equation of state for the anyons occupying the lowest Landau level using the exact ground state basis of the aforementioned Hamiltonian \cite{ouvry3}.

Yet another direction of study is to neglect the mobility of the anyons and consider a system of static interacting anyons. This approach is similar to the study of magnetism, where one only considers static spins that interact with each other and thereby can change their directions \cite{magGr}. Using a generalized Jordan-Wigner construction one can build anyonic oscillators on a 2D square lattice \cite{lerda1993anyons}, which explains the relations of such systems with   quantum groups and $q-$deformations of classical Lie algebras. In a similar vein, it is possible to construct chain or ladder models that are equipped with anyonic degrees of freedom and investigate the properties of these systems \cite{trebst, gils1, gils2, poilblanc1, poilblanc2, poilblanc3, poilblanc4, poilblanc5, gils3}.

Here, we show that anyons can also form anyonic quasiparticles and that the emergent anyons have different braiding properties than the original anyons. As a model system, we consider states that are related to the family of Laughlin states on a lattice in the plane \cite{latticeLaughlin}. Considering the systems on a lattice simplifies the numerical computations and allows us to study both hole type and particle type quasiparticles. One can approach the continuum limit by increasing the number of lattice sites, and in this limit the wavefunction coincides with one of the states in the lowest Landau level ground state basis of the many-anyon continuum Hamiltonian studied in \cite{ouvry1}. The Laughlin states are characterized by the Landau level filling factor $1/q$. If $q$ is odd, the states describe fermions, if $q$ is even, they describe bosons, and if $q$ is non-integer, they describe anyons. We show how hole type and particle type anyonic quasiparticles can be added to the states with non-integer $q$, and we compute the density profiles and braiding properties of the emergent anyons.

The paper is structured as follows. In Sec.\ \ref{sec:model}, we introduce the anyonic wavefunctions that we investigate in this work and explain how they are related to anyonic wavefunctions in the continuum. In Sec.\ \ref{sec:density}, we modify the states to add anyonic quasiparticles, and we compute the density profiles of the emergent anyons. In Sec.\ \ref{sec:braid}, we confirm the anyonic nature of the emergent anyons by computing their braiding statistics. Section \ref{sec:conclusion} concludes the paper.

\section{Model}\label{sec:model}

Our starting point is a family of Laughlin states on a lattice in the two-dimensional complex plane \cite{latticeLaughlin}. The lattice points are at the positions $(\mathrm{Re}(z_j),\mathrm{Im}(z_j))$, $j\in\{1,2,\ldots,N\}$. For simplicity, we choose a square lattice, and in analogy to a fractional quantum Hall droplet, we choose the boundary of the lattice to be roughly circular by only considering the lattice sites inside a circle of radius $R$. We use $n_j$ to denote the number of particles on the $j$th site, and each site can be either empty ($n_j = 0$) or occupied by one particle ($n_j = 1$). The considered states take the form
\begin{equation}\label{StateVec}
|\psi_{q}^{\eta}\rangle=
\sum_{n_1,n_2,\ldots,n_N}\psi_{q}^{\eta}\;|n_1,n_2,\ldots,n_N\rangle,
\end{equation}
where
\begin{equation}\label{state}
\psi_q^\eta=
\mathcal{C}^{-1}\;\delta_{n}
\;\prod_{i<j}(z_{i}-z_{j})^{qn_{i}n_{j}}
\;\prod_{i\neq j}(z_{i}-z_{j})^{-\eta n_{i}}.
\end{equation}
Here, $q$ is the number of flux quanta per particle, $\eta$ is the number of flux quanta per lattice site, $\mathcal{C}$ is a real normalization constant, and $\delta_n=1$ if there are
\begin{equation}\label{par}
\sum_in_i=\frac{N\eta}{q}
\end{equation}
particles in the system and $\delta_n=0$ otherwise. Note that $N\eta$ is the total magnetic flux, so that \eqref{par} precisely expresses that there are $q$ flux quanta per particle. The exponents appearing in \eqref{state} need not be integer, and to fully define the states, we therefore need to specify the branch used. We do this as explained in Fig.\ \ref{LatticeOrdering}.

For integer $q$, the states \eqref{state} are the Laughlin states of fermions or bosons except that both the particles and the background charge are restricted to be on the specified lattice sites \cite{latticeLaughlin}. We can also write the states in the alternative form
\begin{equation}\label{StateAlt}
\psi_q^\eta\propto\delta_{n}\;\prod_{i<j}(Z_{i}-Z_{j})^{q}\;
\prod_{\{i,j|Z_i\neq z_j\}}(Z_{i}-z_{j})^{-\eta},
\end{equation}
where $Z_j\in\{z_1,z_2,\ldots,z_N\}$ is the position of the $j$th particle. The choice of branch is the same as before, and we number the particles such that particles with higher indices are on sites with higher indices (i.e., if $Z_j=z_n$, $Z_k=z_m$, and $n<m$, then $j<k$). From this expression, we observe that the wavefunction acquires a phase factor $e^{\pi i q}$ if the $i$th particle and the $j$th particle are exchanged
in the counterclockwise direction (see Appendix~\ref{Appendix: Exchange} for details). The state hence describes fermions if $q$ is odd, hardcore bosons if $q$ is even, and hardcore anyons if $q$ is non-integer.

Note that we can increase the number of lattice sites in the system, while keeping the total flux $N\eta$ and the number of particles $\sum_i n_i$ constant, by choosing $\eta$ to be inversely proportional to $N$. Thus the parameter $\eta$ is a handle for interpolating between, e.g., the lattice with one flux unit per site ($\eta = 1$) and the continuum limit ($\eta \to 0^+$). In the continuum limit, the states \eqref{StateAlt} become
\begin{equation}\label{StateContExact}
\psi_q^\textrm{cont}\propto\delta_{n}\;\prod_{i<j}(Z_{i}-Z_{j})^{q}\;
\exp \left(-\sum_{i = 1}^N \frac{\left|Z_{i}\right|^{2}}{4}\right),
\end{equation}
where $q$ is a positive real number and we have discarded some single particle phase factors that do not modify the topological properties of the states.

The state \eqref{StateContExact} is within the lowest Landau level ground state basis of the many-anyon continuum Hamiltonian studied in \cite{ouvry1}. Specifically, it is the state in Eq.\ (26) of Ref.\ \cite{ouvry1} with the following identifications:
\begin{gather}
\alpha = -q, \nonumber \\
\omega_{c} = 1/2, \\
l_{1} = l_{2} = \cdots = l_{N} = 0. \nonumber
\end{gather}
Let us also point out that one can trivially construct a Hamiltonian for the many-anyon wavefunction [\eqref{StateVec} with non-integer $q$] as follows:
\begin{equation}
H = \mathbb{1} - |\psi_{q}^{\eta}\rangle \langle \psi_{q}^{\eta}|.
\end{equation}
One can hence physically realize the lattice wavefunction by realizing this Hamiltonian.

\section{Density difference profiles of the emergent anyons}\label{sec:density}

\begin{figure*}
\includegraphics[width=0.33\textwidth]{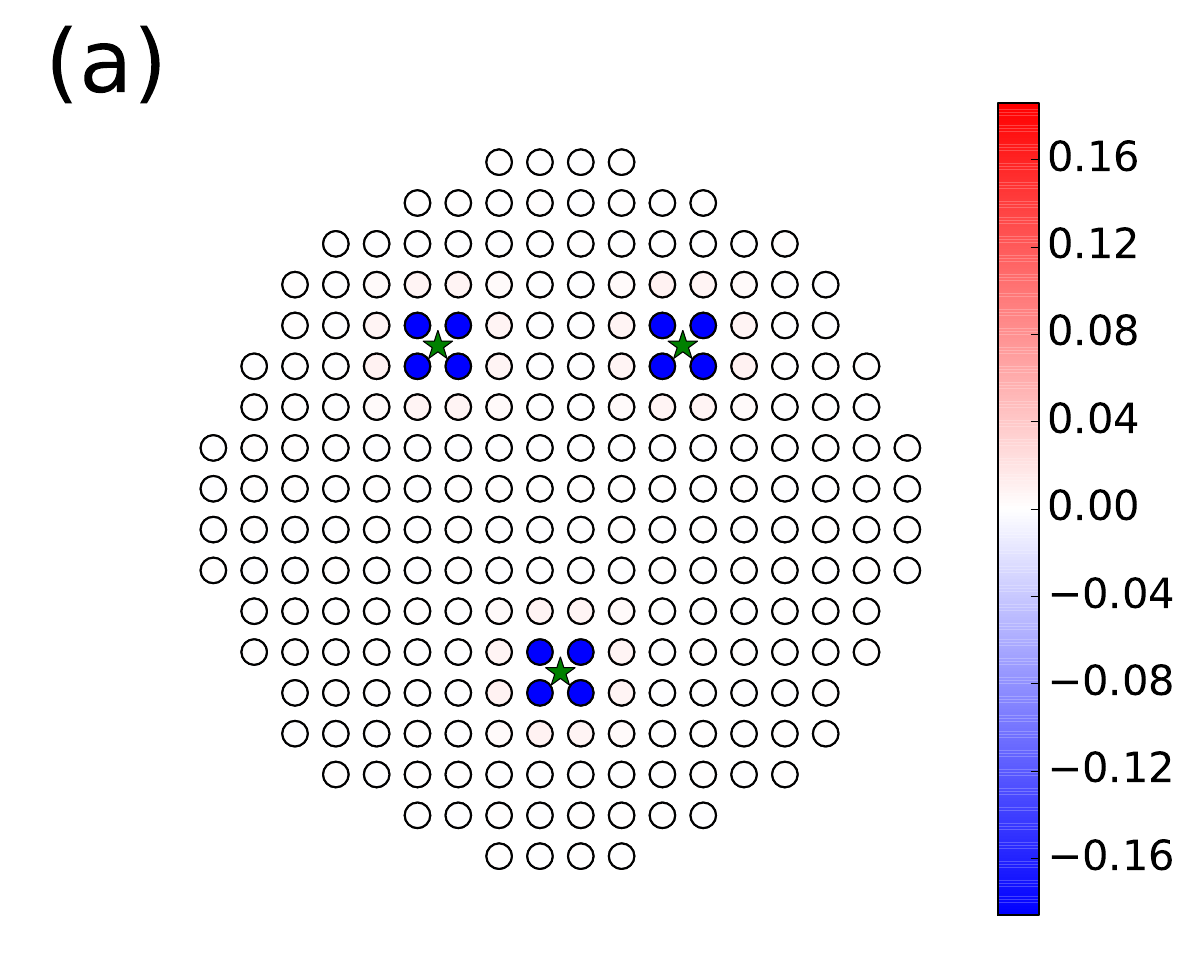}\hfill
\includegraphics[width=0.33\textwidth]{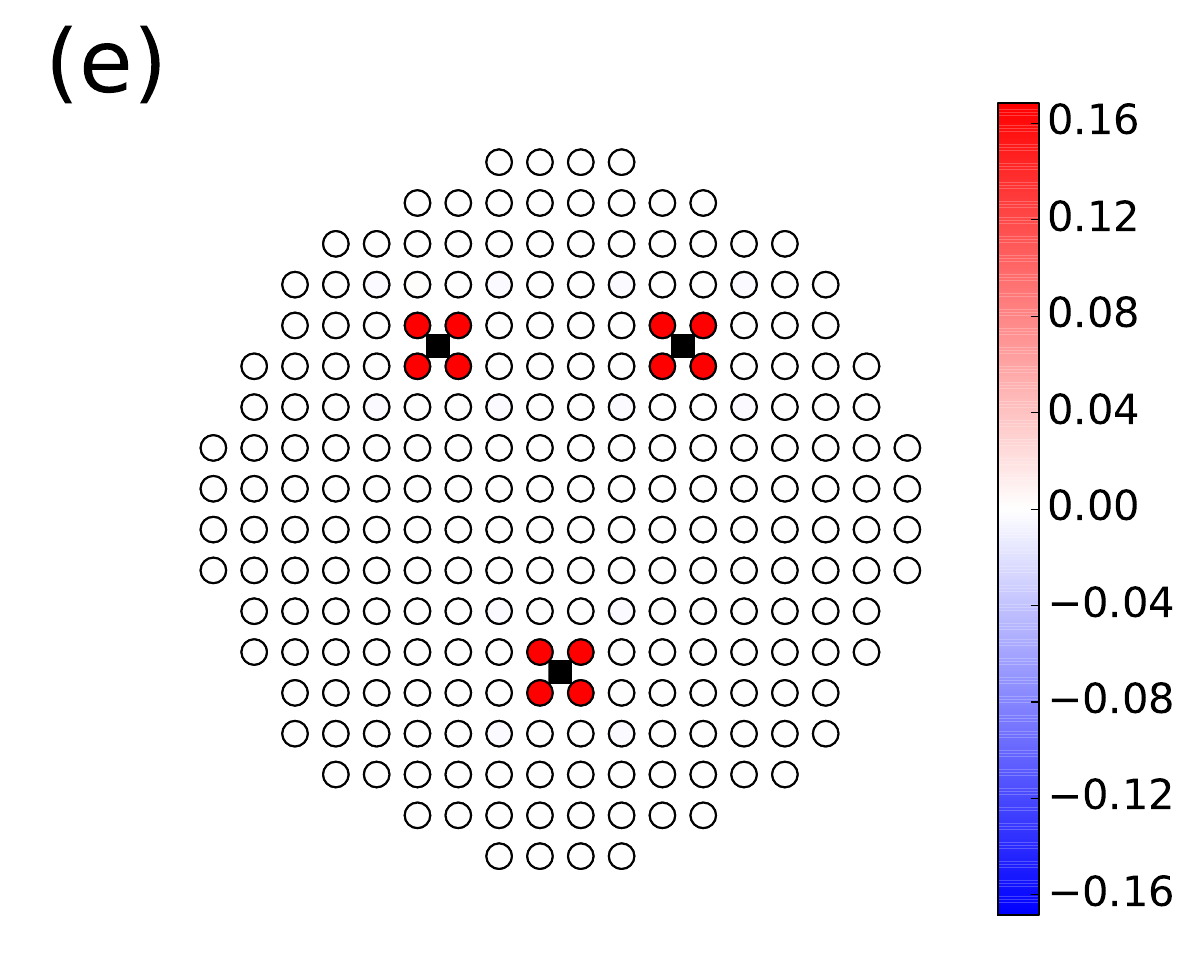}\hfill
\includegraphics[width=0.33\textwidth]{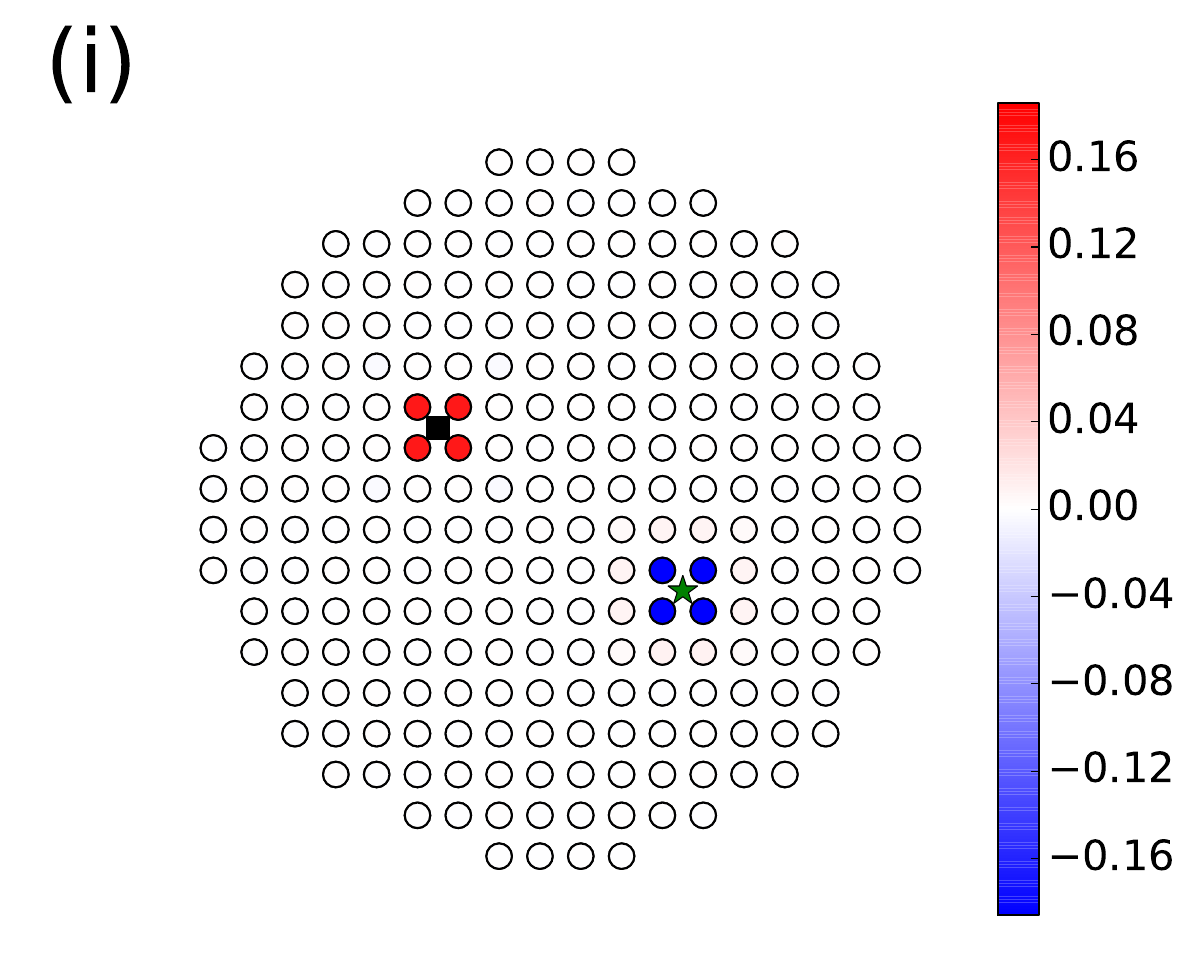}\\
\includegraphics[width=0.33\textwidth]{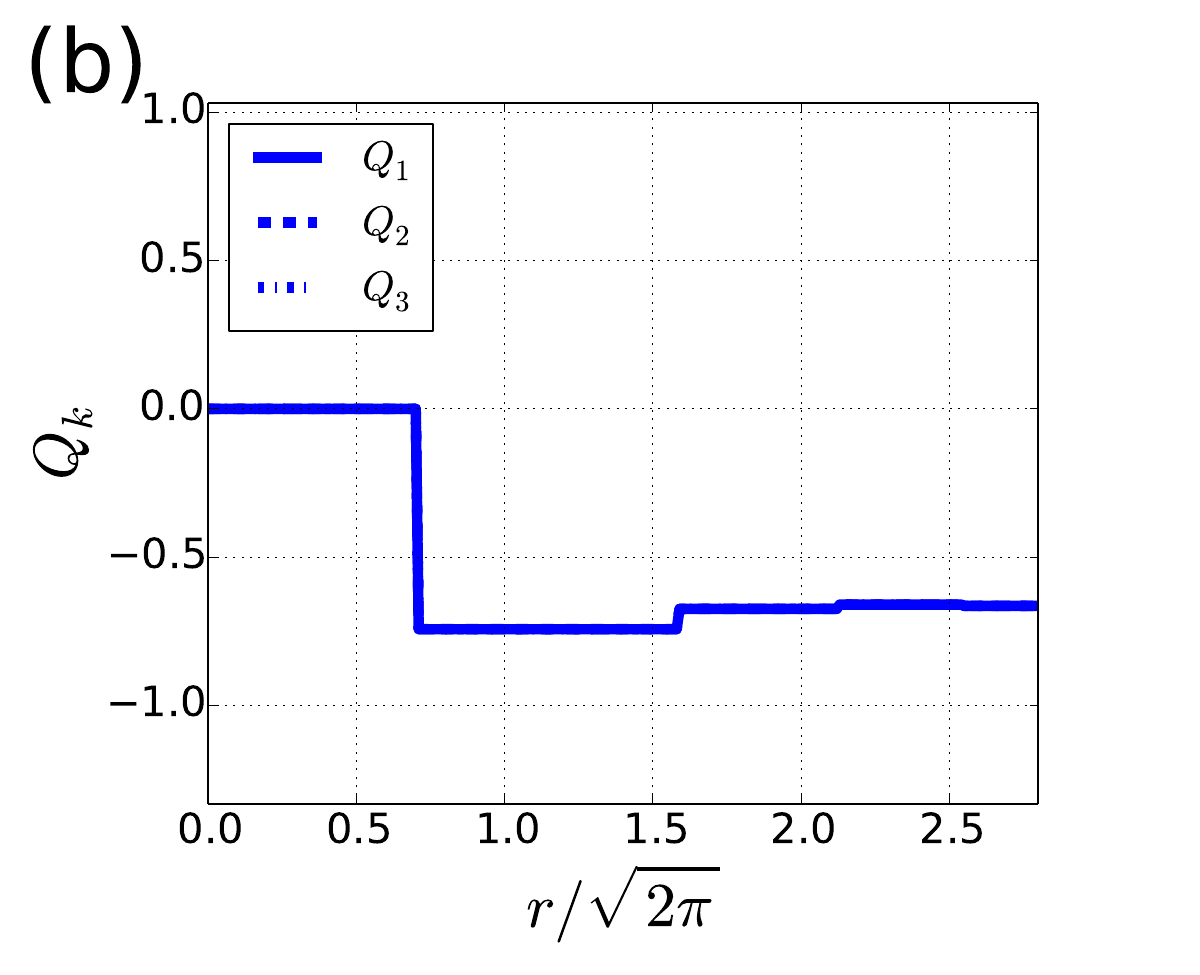}\hfill
\includegraphics[width=0.33\textwidth]{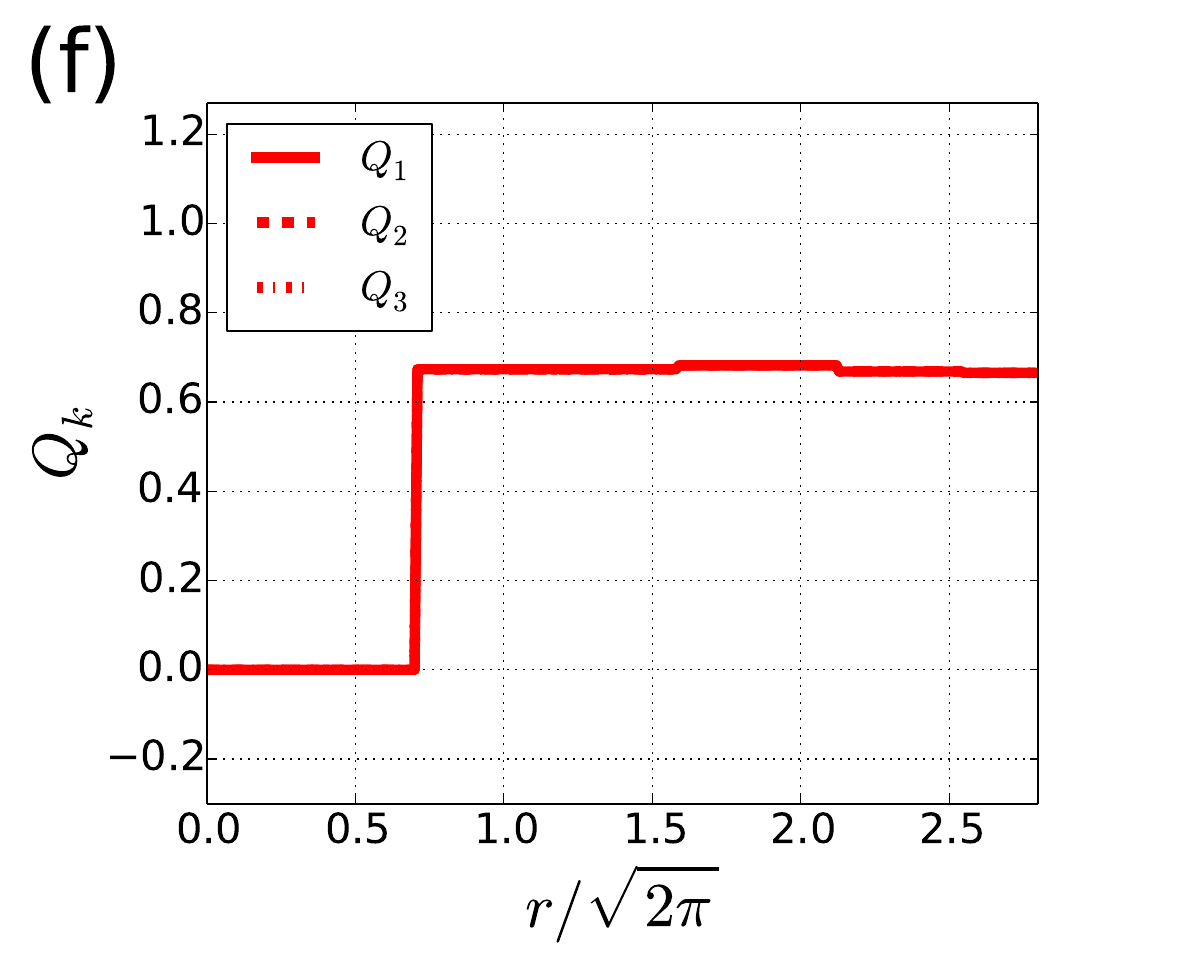}\hfill
\includegraphics[width=0.33\textwidth]{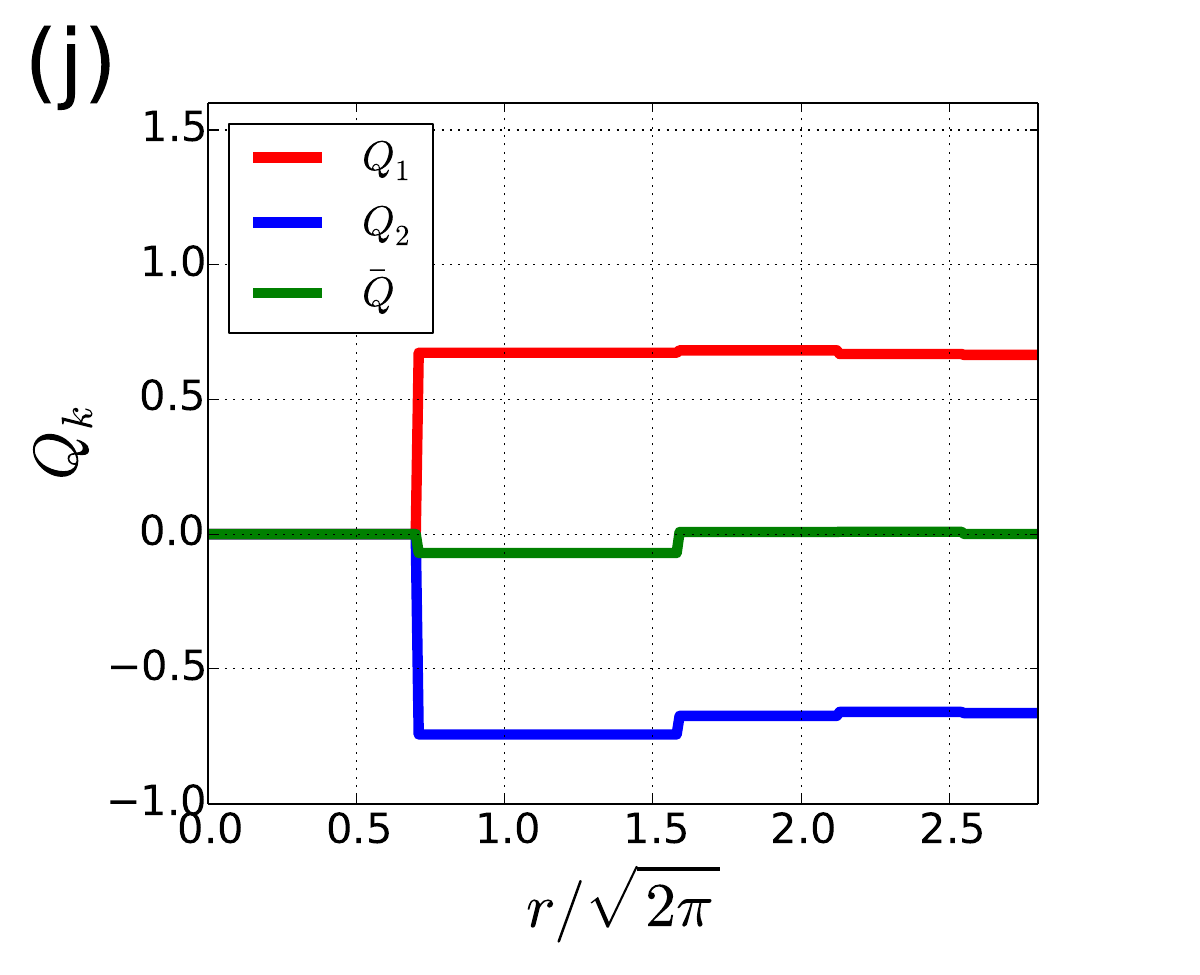}\\
\includegraphics[width=0.33\textwidth]{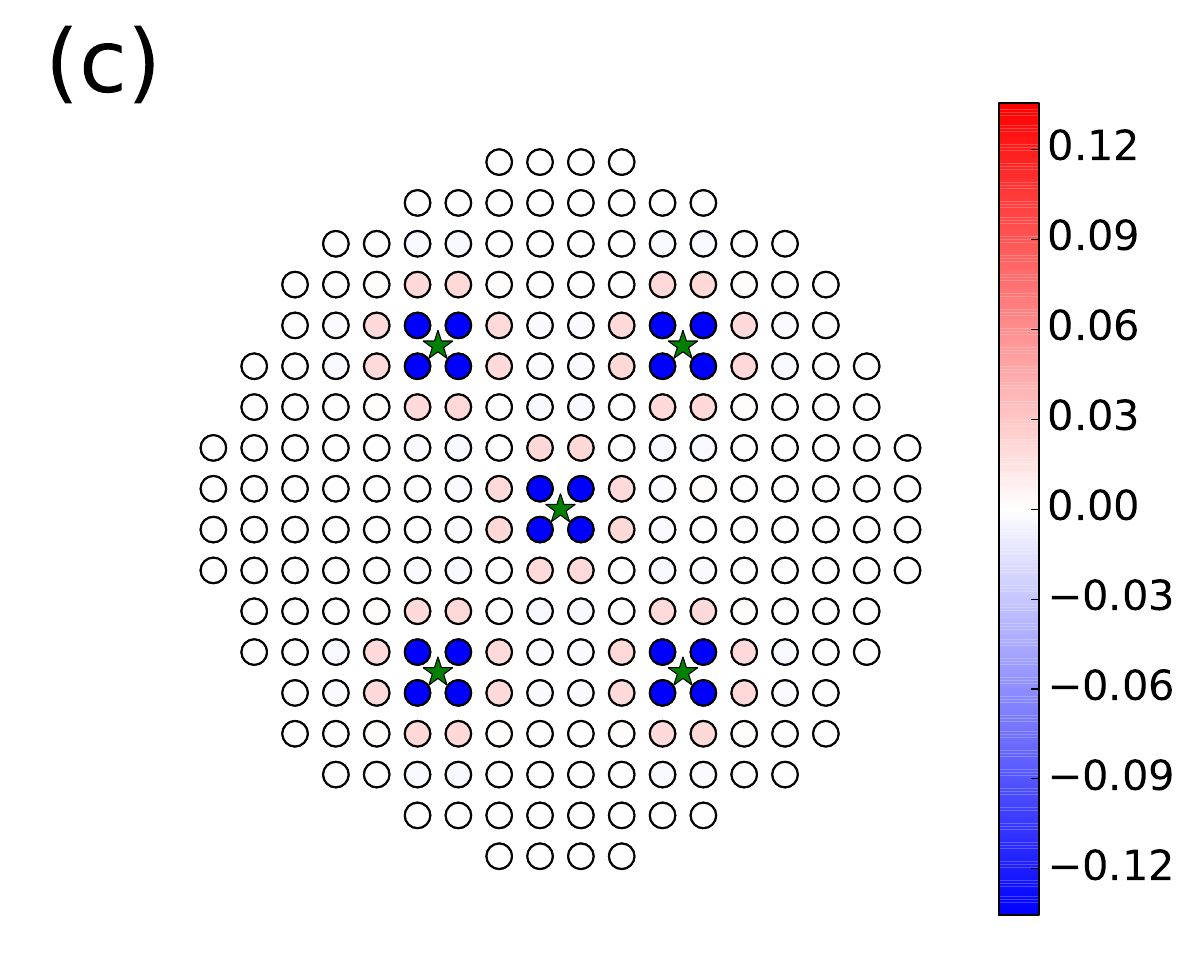}\hfill
\includegraphics[width=0.33\textwidth]{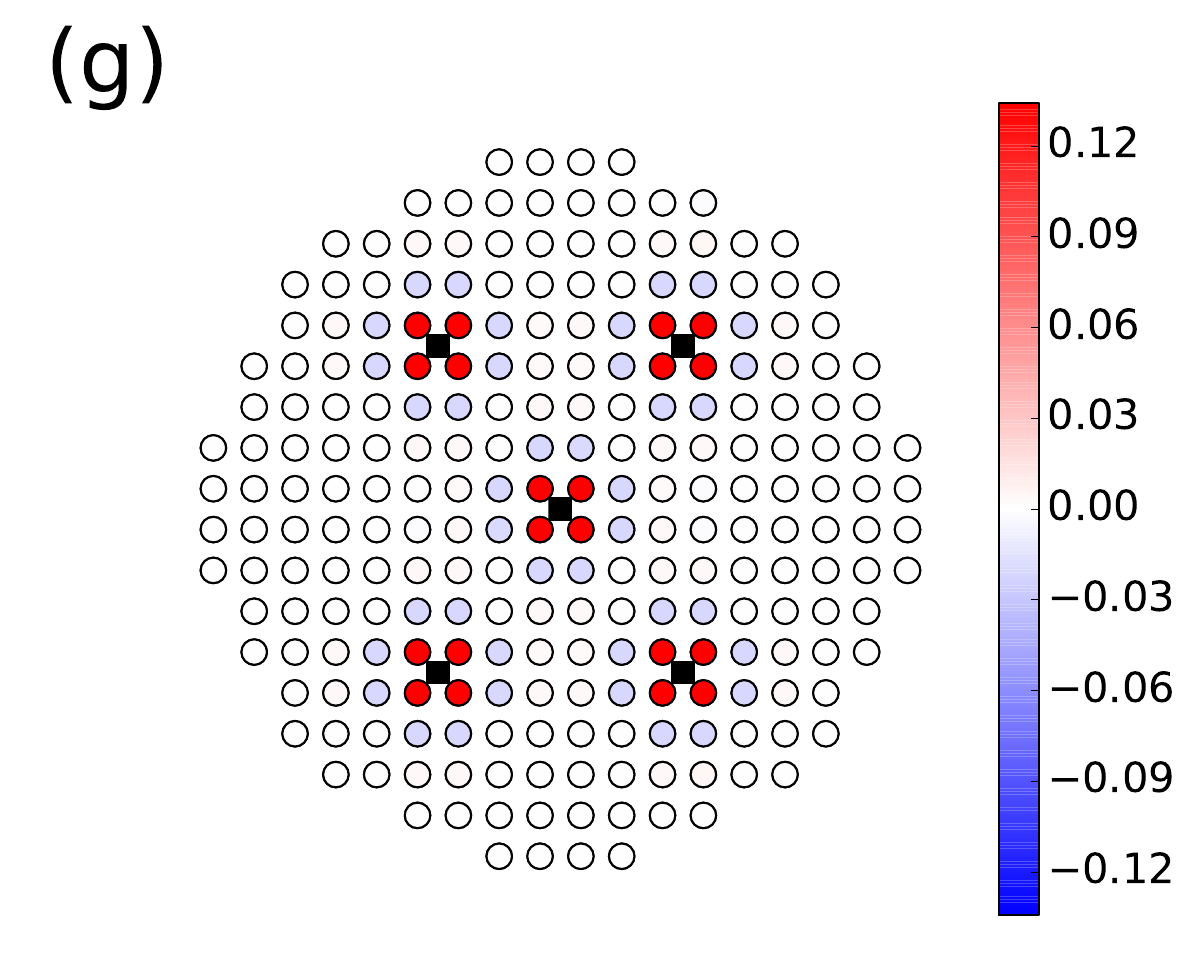}\hfill
\includegraphics[width=0.33\textwidth]{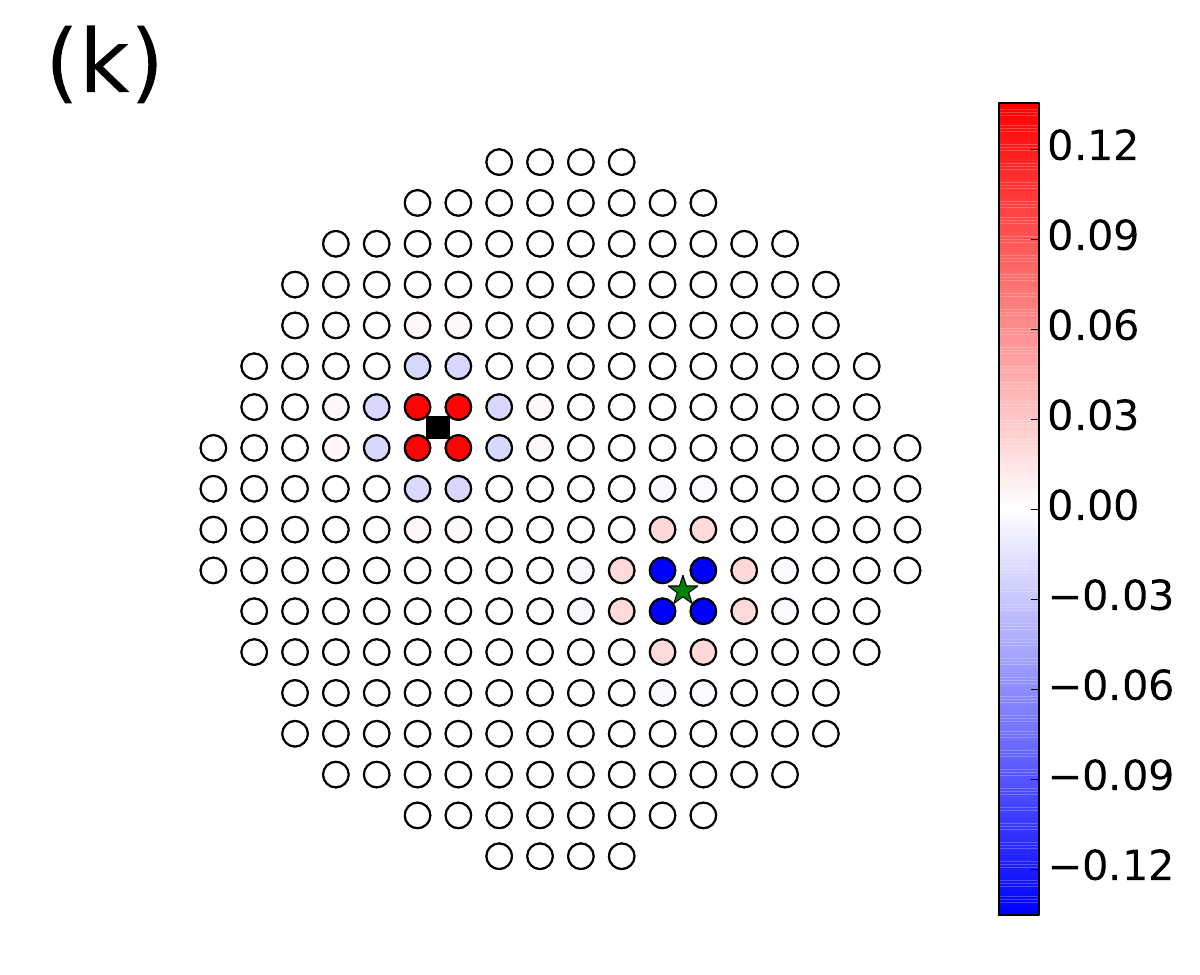}\\
\includegraphics[width=0.33\textwidth]{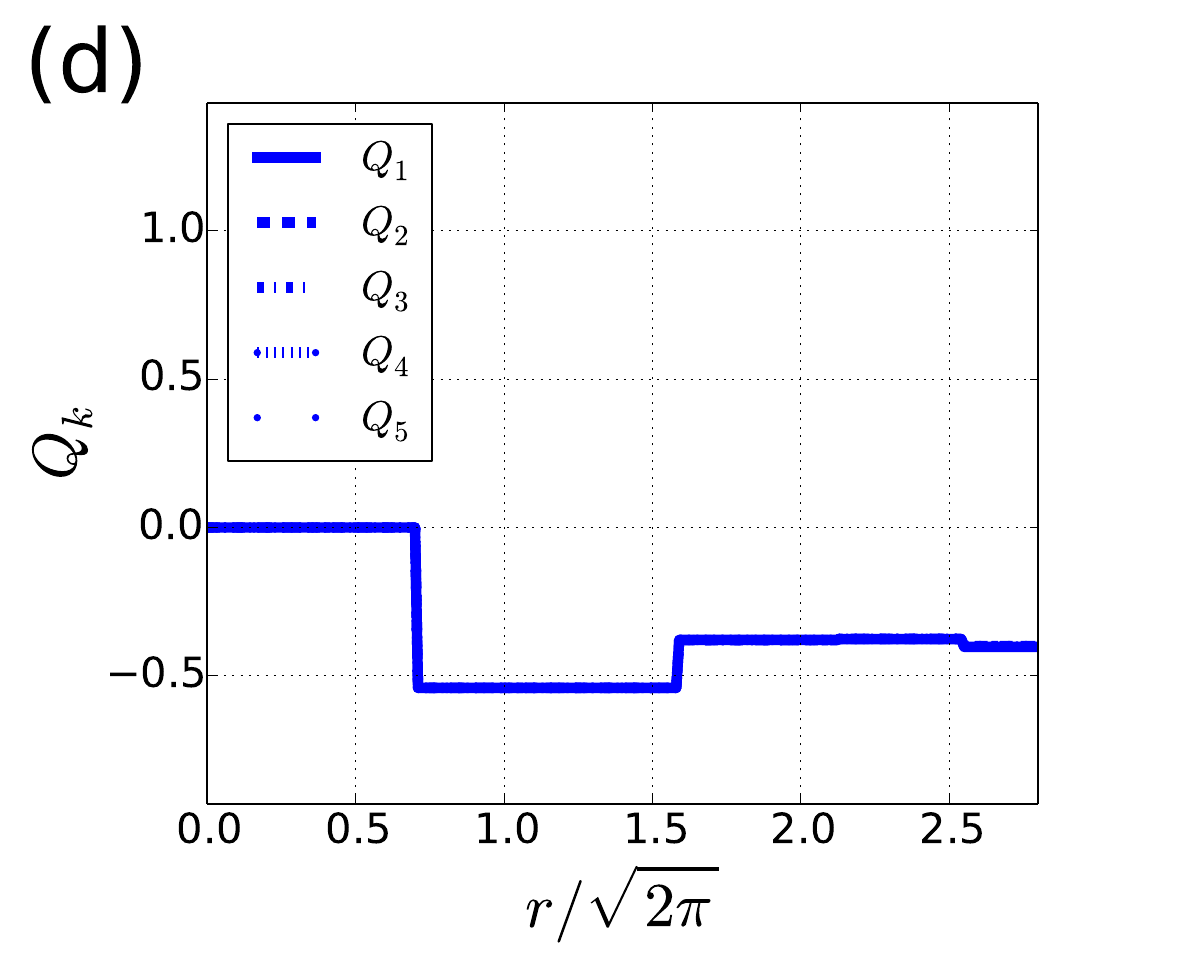}\hfill
\includegraphics[width=0.33\textwidth]{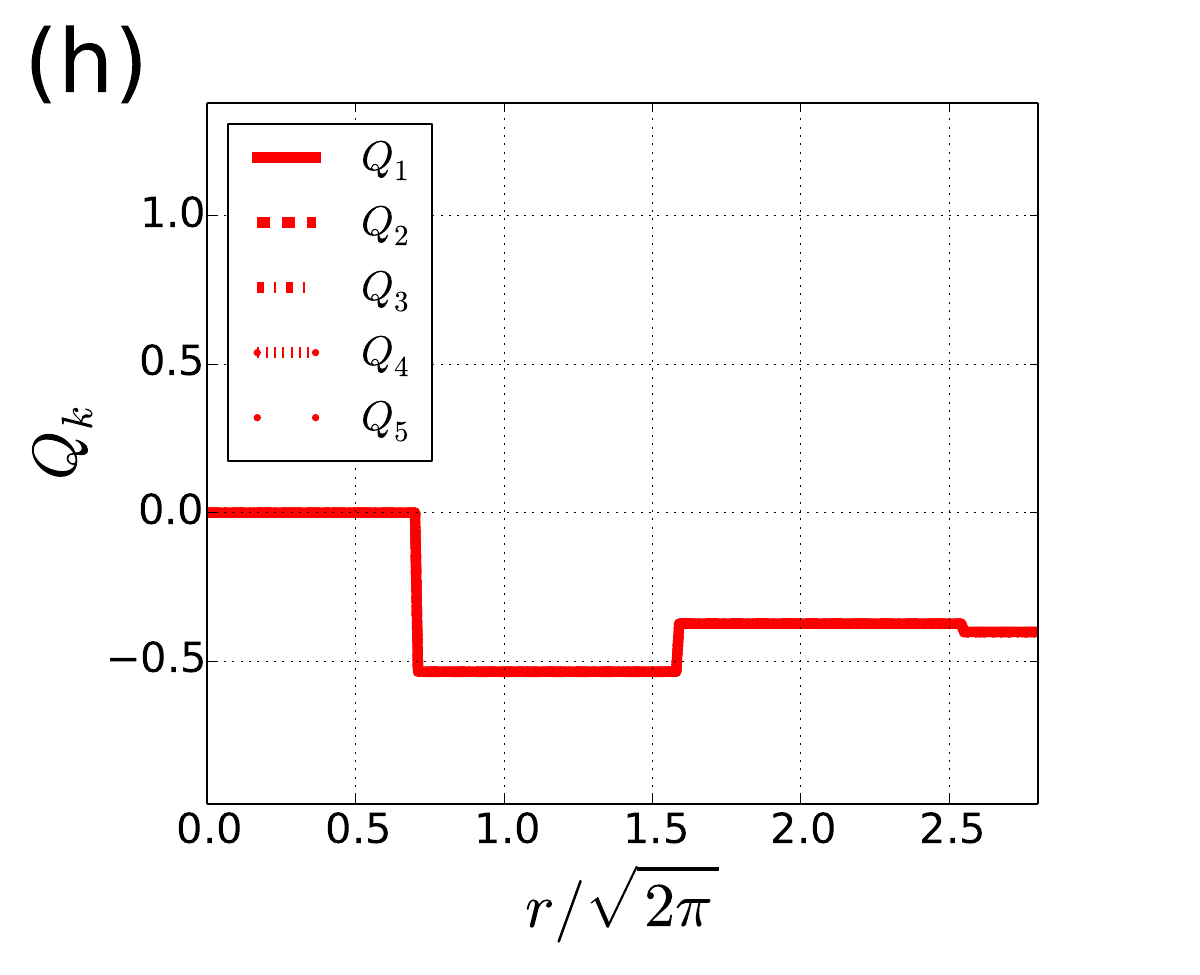}\hfill
\includegraphics[width=0.33\textwidth]{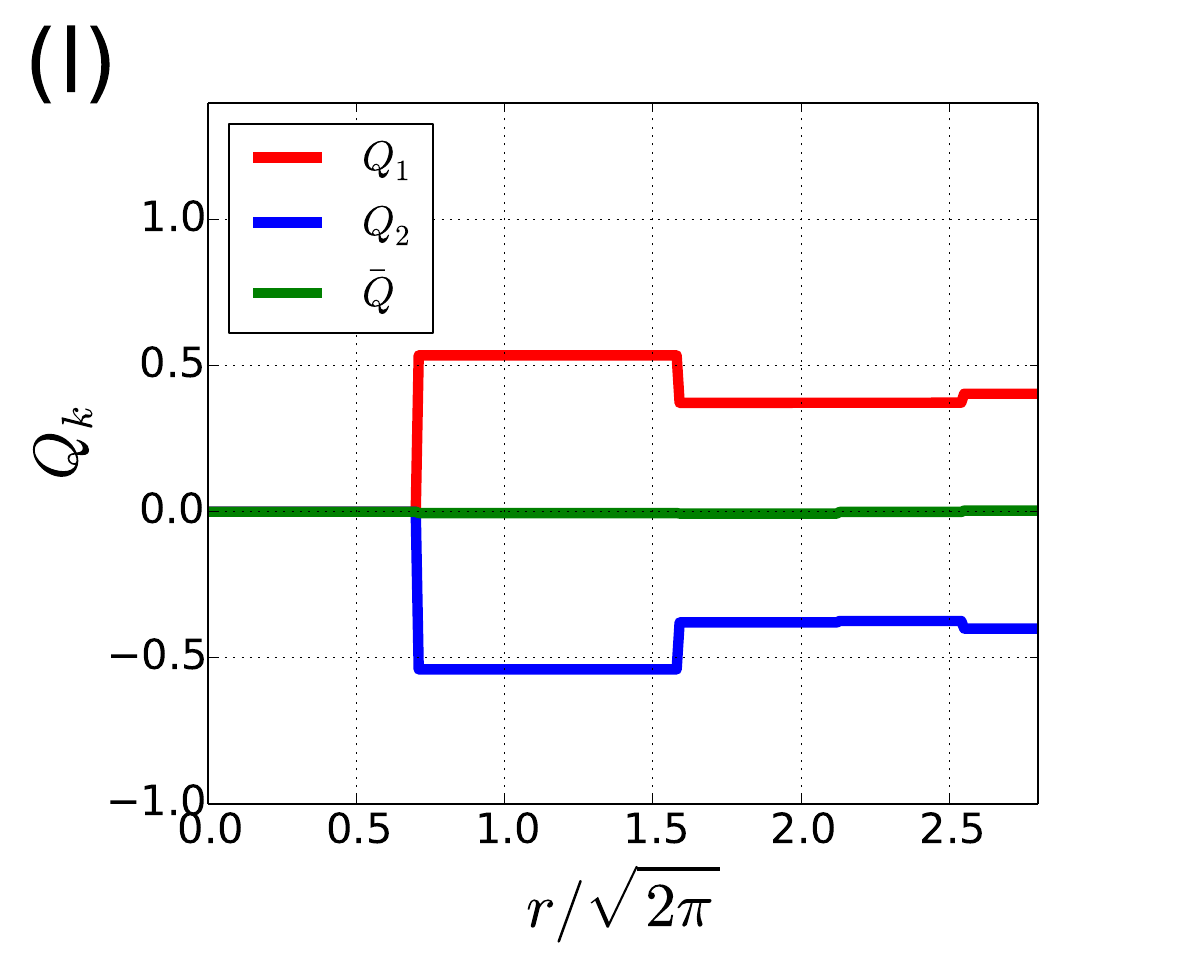}
\caption{(a),(c),(e),(g),(i),(k): Circles represent lattice sites, stars represent hole type quasiparticles, and squares represent particle type quasiparticles. The number of lattice sites is $N = 240$ and we take $q=3/2$ in (a), (e), (i) and $q=5/2$ in (c), (g), (k). We consider the case of one flux unit per lattice site -- i.e., $\eta = 1$. The particle density difference $\rho(z_i)$, defined in \eqref{Density_Profile} to be the difference between the expectation value of the number of particles on the $i$th lattice site with and without quasiparticles present in the system, is plotted with color-bar for the cases of (a) three hole type quasiparticles, (c) five hole type quasiparticles, (e) three particle type quasiparticles, (g) five particle type quasiparticles, and (i),(k) the combination of one hole type and one particle type quasiparticle, respectively. (b),(d),(f),(h),(j),(l): The excess particle numbers, computed from \eqref{Excess_Charge}, are plotted as a function of the radial distance $r$ from the quasiparticle positions. We take $q=3/2$ in (b), (f), (j) and $q=5/2$ in (d), (h), (l). We find that the excess particle numbers approach $\pm 2/3$ in (b), (f), (j) and $\pm 2/5$ in (d), (h), (l) for large $r$. The particle density differences for the hole type quasiparticles and for the particle type quasiparticles are seen to be similar except for the sign, which can be seen from the quantity $\overline{Q} = Q_1 + Q_2$ in (j) and (l). There are small errorbars of order $10^{-4}$ on the data due to the Monte Carlo simulation.}
\label{fig:dpqh1}
\end{figure*}

\begin{figure}
\includegraphics[width=\columnwidth]{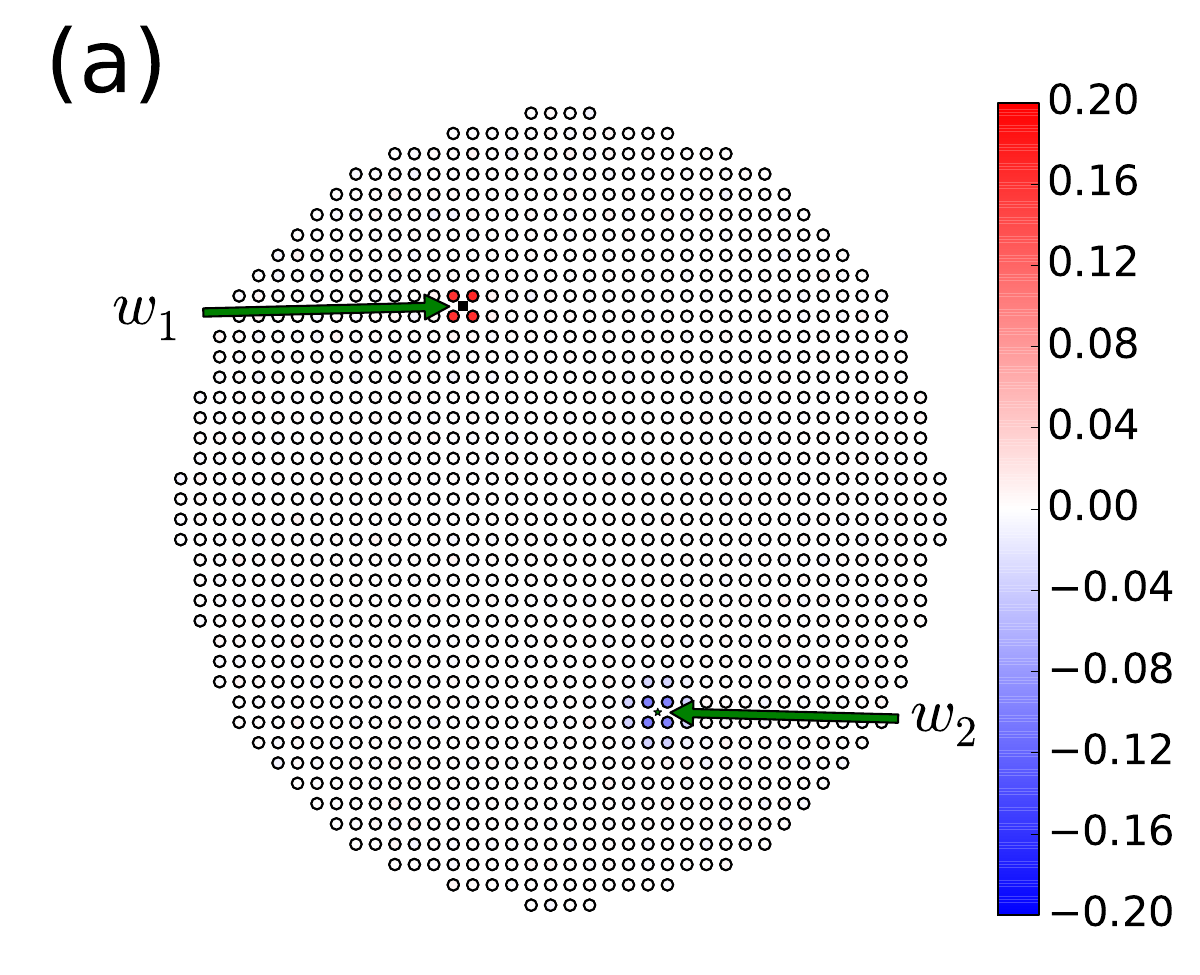}\\
\includegraphics[width=\columnwidth]{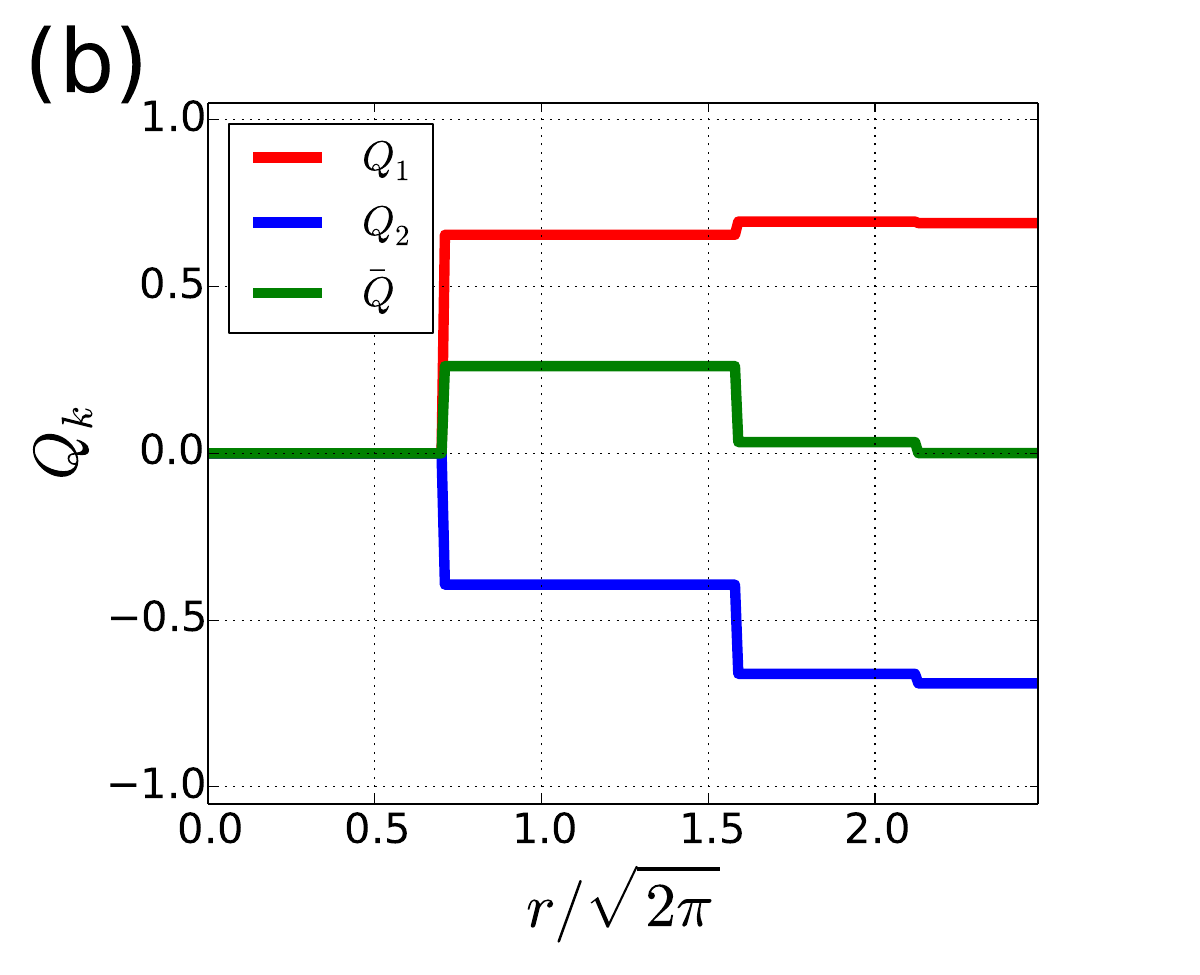}
\caption{(a) The particle density difference $\rho(z_i)$ and (b) the excess particle numbers $Q_k(r)$, as well as $\bar{Q}=Q_1+Q_2$, as a function of the radial distance from the quasiparticle positions for $q=3/2$, $\eta = 30/151$, and $N = 1208$. In (a), the circles represent the lattice sites. The square at $w_1$ and the star at $w_2$ denote a particle type and a hole type quasiparticle, respectively. The system has $N\eta/q = 160$ particles. The plots show that the anyons are screened.}
\label{fig:dpqh15}
\end{figure}

For the bosonic and fermionic lattice Laughlin states with integer $q$, one can add $Q$ anyons at the positions $w_j$ by modifying the wavefunction \eqref{state} into \cite{Anne2}
\begin{multline}\label{anyonstate}
\psi_{q,\vec{w}}^\eta=\mathcal{C}_{\vec{w}}^{-1}\; \delta_{n}\;
\prod_{i,j}(w_i-z_j)^{p_in_j}\;
\prod_{i<j}(z_{i}-z_{j})^{qn_{i}n_{j}}\\
\times\prod_{i\neq j}(z_{i}-z_{j})^{-n_{i}\eta}.
\end{multline}
Here, $\mathcal{C}_{\vec{w}}$ is a real normalization constant and $\vec{w}=(w_1,w_2,\ldots,w_Q)$. The integer $p_j$ is positive if the anyon at $w_j$ is obtained by fusing $p_j$ basic hole type quasiparticles and negative if the anyon at $w_j$ is obtained by fusing $-p_j$ basic particle type quasiparticles. The delta function $\delta_n$ is unity for
\begin{equation}\label{par1}
\sum_{j=1}^Nn_j=\frac{N\eta-\sum_{j=1}^Q p_j}{q}
\end{equation}
and zero otherwise. Equation \eqref{par1} shows that an anyon with positive $p_j$ takes up $p_j$ flux quanta, and the presence of the anyon reduces the number of particles by $p_j/q$. The anyon hence produces a hole corresponding to $p_j/q$ particles. If $p_j$ is negative, the anyon instead creates a region with $-p_j/q$ particles extra. In the following, we will show that the states \eqref{anyonstate} can host anyonic quasiparticles also for cases where $q$ is not an integer. The computations involve both analytical and numerical parts. For the analytical parts, we consider general $q$, and for the numerical parts, we consider the cases $q=3/2$ and $q=5/2$.

As the area per lattice site is the same for all sites in the square lattice, we shall refer to $\langle n(z_i) \rangle \equiv \langle \Phi | n_i | \Phi \rangle$ as the particle density, where $\Phi$ is the state of the system. Similarly, we define the particle density difference
\begin{equation}\label{Density_Profile}
\rho(z_i)=\langle \psi_{q,\vec{w}}^\eta|  n_i|\psi_{q,\vec{w}}^\eta \rangle
-\langle \psi_{q}^\eta | n_i | \psi_{q}^\eta \rangle
\end{equation}
to be the difference between the particle density, when there are quasiparticles in the system, and the particle density, when there are no quasiparticles in the system. The particle density difference describes the profile of the emergent anyons, and if the anyons are properly screened, it is nonzero only in a small region around each $w_k$.

To quantify the number of extra particles in a region around each $w_k$, we define the excess particle number
\begin{equation}\label{Excess_Charge}
Q_k(r) = \sum_{i=1}^N \Theta(r-|z_i-w_k|) \rho(z_i).
\end{equation}
In this expression, we sum $\rho(z_i)$ over a circular region of radius $r$ around $w_k$ as ensured by the Heaviside step function $\Theta$. If the anyons are properly screened, the excess particle number of the $k$th anyon converges to $-p_k/q$ for large $r$, provided the circular region is away from the edge and away from all other emergent anyons in the system. The quantity $-p_k/q$ is negative for hole type quasiparticles and positive for particle type quasiparticles.

We show numerical results for $\rho(z_i)$ and $Q_k(r)$ in Fig.\ \ref{fig:dpqh1}, where we take $q=3/2$ and $q=5/2$ for $\eta = 1$ and $N = 240$ lattice sites. Fig.\ \ref{fig:dpqh1}(a), (c), (e), (g), and (i), (k) display the particle density differences for the systems containing three hole type quasiparticles, five hole type quasiparticles, three particle type quasiparticles, five particle type quasiparticles, and the combination of one hole type and one particle type quasiparticle, respectively. We observe that $\rho(z_i)$ is only nonzero in a small region around each $w_k$. The excess particle numbers of the anyons are plotted as a function of the radial distance from the anyons in Fig.\ \ref{fig:dpqh1}(b), (d), (f), (h), and (j), (l), and it is observed that the excess particle number approaches $p_k/q \sim \pm 2/3$ in (b), (f), (j) and $p_k/q \sim \pm 2/5$ in (d), (h), (l), respectively. These observations show that the anyons are screened, and that the anyons are a few lattice spacings wide.

In Fig.\ \ref{fig:dpqh15}, we show the particle density difference for the combination of one hole type and one particle type quasiparticle, when we are much closer to the continuum limit. We take $q = 3/2$, $\eta = 30/151$, and $N = 1208$. Also in this case we observe screening. This property is retained for even smaller values of $\eta$. Proceeding this way, one can obtain a consistent continuum limit for the hole type quasiparticle. However, such a limit generally does not exist for the particle type quasiparticle \cite{ContQE}.

In Figs.\ \ref{fig:dpqh1} and \ref{fig:dpqh15}, we chose the quasiparticle positions $w_{i}$ to always coincide with centers of the square plaquettes. In general, Eq.\ \eqref{anyonstate} produces screened quasiparticles for any locations. This is true, even when we consider a particle type quasiparticle $(p_{i} = -1)$ arbitrarily close to a lattice position $z_{j}$. In this case, $n_{j} \rightarrow 1$. The wavefunction, however, remains normalizable since the infinite factor $\left(w_{i} - z_{j}\right)^{-1}$ is a constant that can be absorbed in the normalization constant, and after this has been done, the normalization constant is finite. In fact, the continuum limit for the particle type quasiparticle, when $q$ is an integer, only exists when the quasiparticle is located on top of a lattice site \cite{ContQE}.

\section{Braiding properties of the emergent anyons}\label{sec:braid}

We now proceed to determine the result of braiding the coordinate $w_k$ around the coordinate $w_j$. These coordinates correspond to the positions of two emergent anyons in \eqref{anyonstate}. When $w_{k}$ is adiabatically moved on a closed contour $c$, the wavefunction changes as $|\psi\rangle \to \mathbb{M}e^{i\theta_{k}}|\psi\rangle$, where $\mathbb{M}$ is the monodromy and $\theta_{k}$ is the Berry phase \cite{berryphase,read08}. Although the ground state wavefunction \eqref{state} is multivalued because the factors $(z_{i} - z_{j})$ have non-integer exponents, the powers $p_{i}n_{j}$ of the terms involving the emergent anyons $w_{i}$ are integers. As a result, the monodromy $\mathbb{M}$ is the identity and only the Berry phase contributes nontrivially to the braiding statistics.

The Berry phase is computed as
\begin{align}
\theta_k&=i\oint_c\langle\psi_{q,\vec{w}}^\eta|\frac{\partial \psi_{q,\vec{w}}^\eta}{\partial w_k}\rangle dw_k+\text{c.c.}\nonumber\\
&=i\frac{p_k}{2}\oint_c\sum_i\frac{\langle\psi_{q,\vec{w}}^\eta| n_i|\psi_{q,\vec{w}}^\eta\rangle}{w_k-z_i}dw_k+\text{c.c.},
\end{align}
where c.c.\ is the complex conjugate of the first term. We are interested in $\Delta\theta_k=\theta_{k,\text{in}}-\theta_{k,\text{out}}$,
where $\theta_{k,\text{in}}$ ($\theta_{k,\text{out}}$) is the Berry phase when $w_j$ is well inside (outside) the closed path $c$.
We have
\begin{equation}\label{DelTheta}
\Delta \theta_k = i\frac{p_k}{2}\oint_c\sum_i\frac{\langle n_i\rangle_{\text{in}}-\langle n_i\rangle_{\text{out}}}{w_k-z_i}dw_k+\text{c.c}.
\end{equation}
We need to remain on the same branch of the anyon wavefunction \eqref{anyonstate} when we calculate $\langle n_i\rangle_{\text{in}}$ and $\langle n_i\rangle_{\text{out}}$ in Eq.\ \eqref{DelTheta} for different $w_{k}$ values on the contour $c$. This is done by retaining the same lattice orderings and the same branch cuts for the individual lattice coordinates throughout, as shown in Fig.\ \ref{LatticeOrdering}. This in turn defines the phases of $(z_{i} - z_{j})$ uniquely. Note that having the lattice on a topologically nontrivial manifold, such as a torus \cite{torusCFT}, introduces additional generators in the underlying braid group \cite{wubraid}. This makes it more difficult to compute the braiding properties on such manifolds.

If the emergent anyons are properly screened, we have that $\langle n_i\rangle_{\text{in}}-\langle n_i\rangle_{\text{out}}$ is nonzero only close to the two possible positions of $w_j$ and is independent of $w_k$. We can then take the factor $\langle n_i\rangle_{\text{in}}-\langle n_i\rangle_{\text{out}}$ outside the integral, which leads to
\begin{equation}
\Delta \theta_k = -2\pi p_k\sum_{i \in {I_c}} \left(\langle n_i\rangle_{\text{in}}-\langle n_i\rangle_{\text{out}}\right),
\end{equation}
where $I_c$ is the set of indices $i$ for which $z_i$ is inside $c$. The sum is the excess particle number for the $j$th anyon, and it follows that
\begin{equation}\label{braidangle}
\Delta\theta_k = 2\pi p_k p_j /q.
\end{equation}
This is the same result as the Berry phase for the Laughlin states, except that $q$ is now a non-integer. This confirms that the emergent anyons indeed have anyonic braiding statistics if the assumption that the emergent anyons are properly screened is true. The numerical results of the previous section show that screening occurs for $q=3/2$ and $q=5/2$, and also when we approach the continuum limit.

\section{Conclusions}\label{sec:conclusion}

We have shown that systems consisting of many anyons can support the formation of anyonic quasiparticles, and that the braiding properties of the emergent anyons can differ from the properties of the original anyons. We have also
shown that Laughlin states with non-integer $q$ provide models of anyons, where these phenomena occur. The considered models are defined on lattices, and a continuum limit, in which the particles can be practically anywhere in the two-dimensional plane, can be approached by increasing the number of lattice sites, while keeping the number of particles and the total magnetic flux fixed. In these systems, the wavefunction changes by the phase factor $e^{2\pi i q}$ if one exchanges two of the original anyons twice in the counterclockwise direction. In the continuum limit, the anyonic wavefunction without emergent quasiparticles maps to a state in the lowest Landau level ground state basis of the many-anyon continuum Hamiltonian studied in \cite{ouvry1}.

For the emergent anyons, we have shown analytically that if they are properly screened, then the wavefunction changes by the phase factor $e^{i2\pi p_k p_j/q}$ if one braids an emergent anyon, consisting of $p_k$ basic hole type quasiparticles (or $-p_k$ basic particle type quasiparticles if $p_k<0$), around another emergent anyon, consisting of $p_j$ basic hole type quasiparticles (or $-p_j$ basic particle type quasiparticles if $p_j<0$). We have shown numerically that the emergent anyons are screened for the lattice models with $q=3/2$ and $q=5/2$, and also when the continuum limit is approached. Given the plasma analogy for the Laughlin states, we expect that this result holds also for other values of non-integer $q$, as long as $q$ is not too large. We have found that the emergent anyons have radii of order a few lattice constants when the number of flux units per lattice site is one.

The obtained results motivate several further investigations. In particular, it would be interesting to see which types of anyons can appear as quasiparticles in systems consisting of various types of non-Abelian anyons. It would also be interesting to describe the ground state degeneracy in such systems. One can use braiding operations to perform unitary transformations within the ground state manifold of a system. For systems with a low density of anyons, it is known that some types of anyons allow for a universal set of unitary operations in the ground state manifold, while others do not. A system of fermions only allows for a change of sign of the wavefunction, but if the fermions form anyonic quasiparticles, more operations are possible. Similarly, the results presented here show that forming Abelian quasiparticles in a system of Abelian anyons increases the number of different phases that one can obtain by doing braiding operations. It would be interesting to understand more generally how emergent anyons affect the possible operations that can be done. Under which conditions is it, e.g., possible to achieve a universal set of operations, although the original anyons do not allow for a universal set?

\appendix

\section{Exchange of Two Anyons}
\label{Appendix: Exchange}

\begin{figure*}
\includegraphics[width=0.70\textwidth]{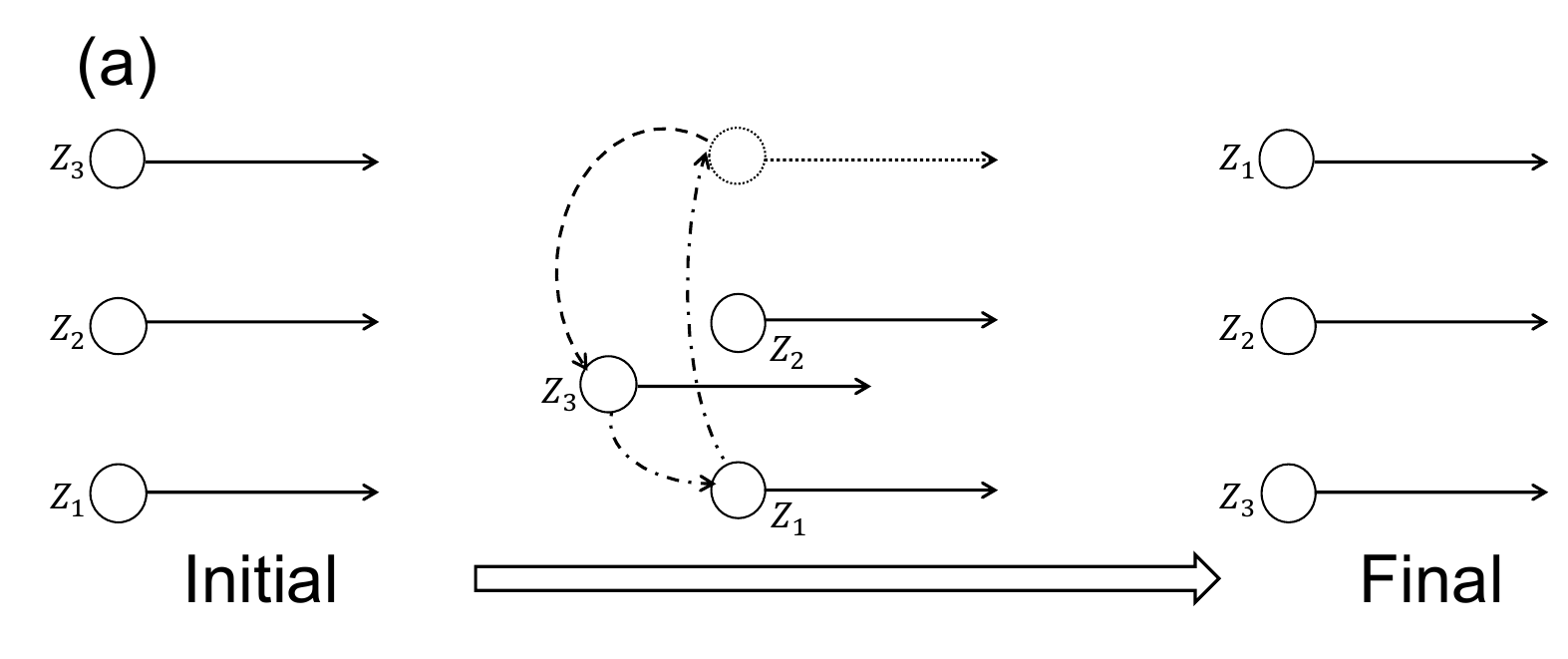}\\
\includegraphics[width=0.70\textwidth]{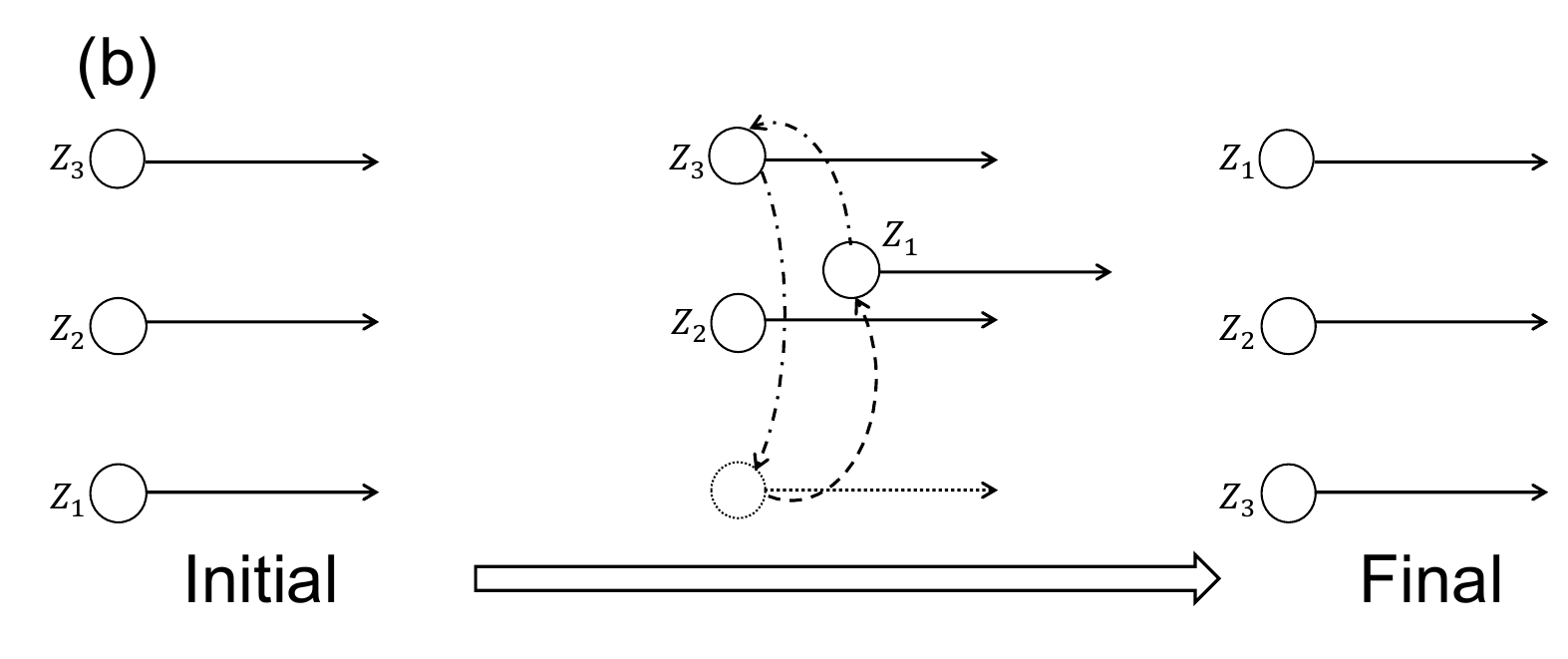}\\
\includegraphics[width=0.70\textwidth]{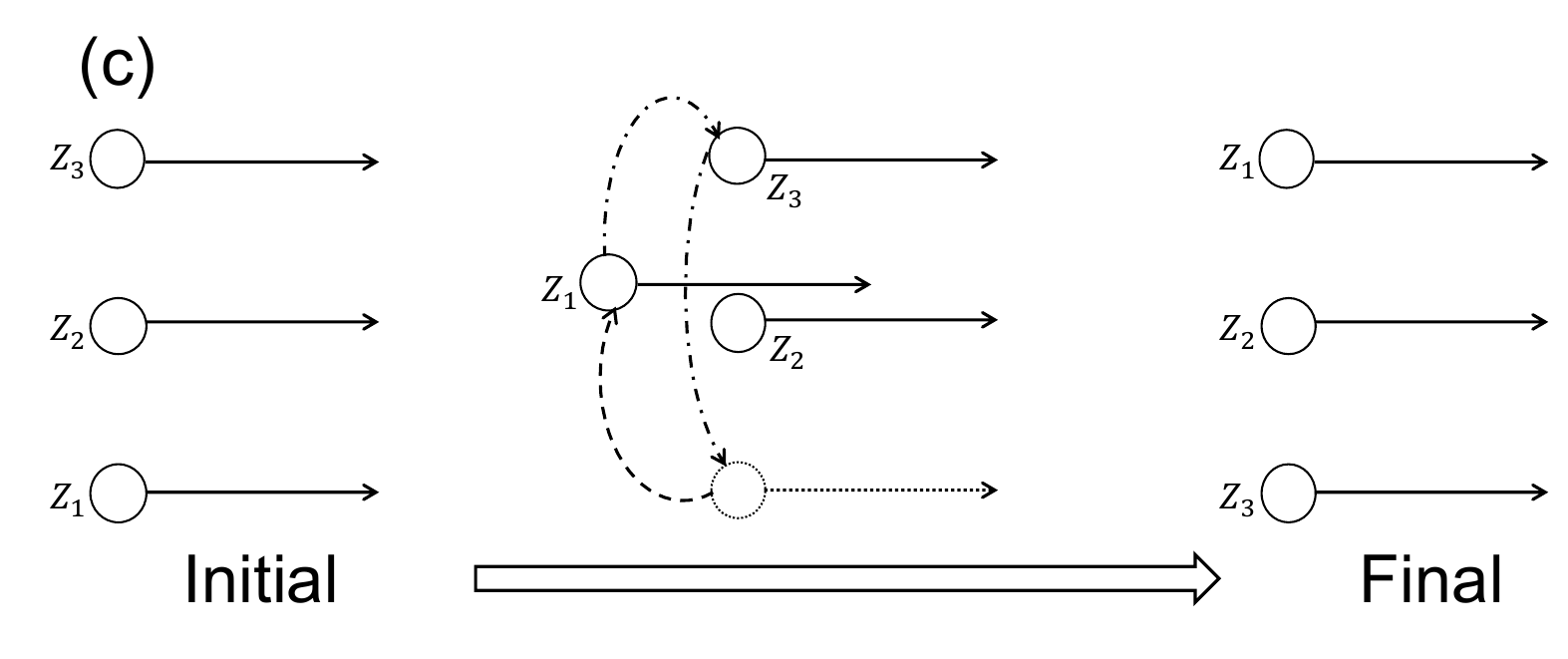}\\
\includegraphics[width=0.70\textwidth]{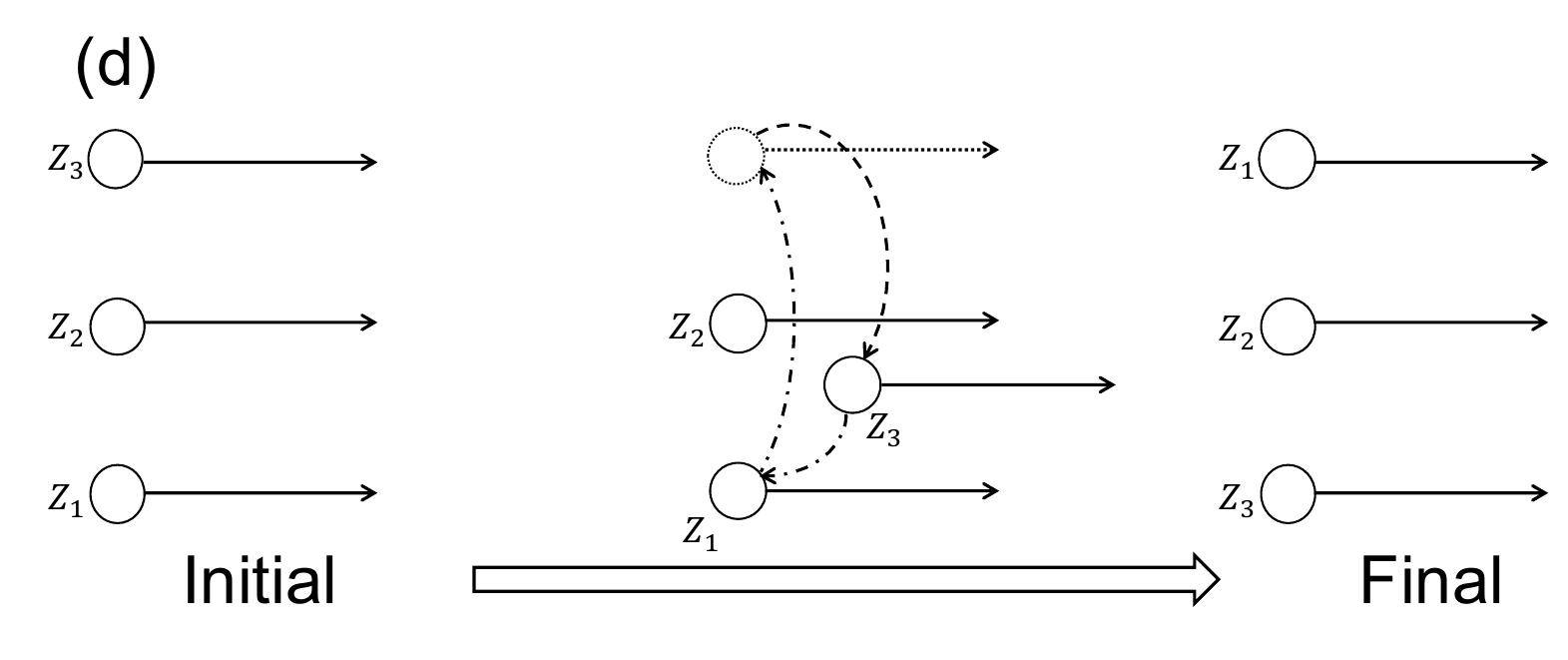}
\caption{Counterclockwise (a,b) and clockwise (c,d) exchanges of the anyons $\#1$ and $\#3$. In exchanges (a) and (c), anyon $\#2$ remains to the right of the exchange loop, whereas it is placed to the left of the loop in (b) and (d). These constitute all the possible ways of exchanging anyons $\#1$ and $\#3$, where anyon $\#2$ is not enclosed in the exchange loop. First, one of the anyons (let us call it A) is adiabatically moved away from the lattice site -- shown by the dashed arrow. In the next step, the other anyon (let us denote it B) is transported to the initial position of anyon A while A moves into the initial position of anyon B. This completes the exchange process. The second step is shown by the dot-dashed arrows.}
\label{fig:phase_jump}
\end{figure*}

\begin{table}[ht]
\begin{center}
\begin{tabular}{| c | c | c | c | c |}
\hline
\bf \makecell{Type of \\ exchange} & \bf \makecell{ Position of \\ anyon $\bf \#2$} & \bf Crossing & $\bf\Phi^{\pm}_{mn}$ & $\bf\Delta \Phi_{mn}$ \\
\hline \hline
\multirow{3}{*}{Counterclockwise} & \multirow{3}{*}{\makecell{Right of \\ the loop}} & $(1 \uparrow, 2 \downarrow)$ & $0$ & $-\pi q$ \\
\cline{3-5} & & $(3 \downarrow, 1 \uparrow)$ & $-2\pi q$ & $+\pi q$ \\
\cline{3-5} & & $(3 \downarrow, 2 \uparrow)$ & $-2\pi q$ & $+\pi q$ \\
\hline \cline{1-5}
\multirow{3}{*}{Counterclockwise} & \multirow{3}{*}{\makecell{Left of \\ the loop}} & $(2 \downarrow, 1 \uparrow)$ & $-2\pi q$ & $+\pi q$ \\
\cline{3-5} & & $(3 \downarrow, 1 \uparrow)$ & $-2\pi q$ & $+\pi q$ \\
\cline{3-5} & & $(2 \uparrow, 3 \downarrow)$ & $0$ & $-\pi q$ \\
\hline \hline
\multirow{3}{*}{Clockwise} & \multirow{3}{*}{\makecell{Right of \\ the loop}} & $(1 \uparrow, 2 \downarrow)$ & $0$ & $-\pi q$ \\
\cline{3-5} & & $(1 \uparrow, 3 \downarrow)$ & $0$ & $-\pi q$ \\
\cline{3-5} & & $(3 \downarrow, 2 \uparrow)$ & $-2\pi q$ & $+\pi q$ \\
\hline \cline{1-5}
\multirow{3}{*}{Clockwise} & \multirow{3}{*}{\makecell{Left of \\ the loop}} & $(2 \downarrow, 1 \uparrow)$ & $-2\pi q$ & $+\pi q$ \\
\cline{3-5} & & $(1 \uparrow, 3 \downarrow)$ & $0$ & $-\pi q$ \\
\cline{3-5} & & $(2 \uparrow, 3 \downarrow)$ & $0$ & $-\pi q$ \\
\hline
\end{tabular}
\end{center}
\caption{The phase discontinuities $\left(\Phi^{\pm}_{mn}\right)$ and the phase changes $\left(\Delta \Phi_{mn}\right)$ in $\left(Z_{m} - Z_{n}\right)^{q}$ as one anyon crosses the other in the counterclockwise and the clockwise exchange processes in Fig.\ \ref{fig:phase_jump}. Here we consider $m<n$ and both $m$ and $n$ are positive integers. The anyon, which is not involved in the exchange process, can remain either to the left or to the right of the exchange loop. This is specified in the second column. In the third column, we describe the different crossings by the notation $(i\uparrow / \downarrow, j\uparrow / \downarrow)$, where the $i$th anyon is on the left, when crossing the $j$th anyon. The directions of motion of the anyons while crossing is shown by either $\uparrow$ or $\downarrow$. In the last column, we list the phase changes in the factors $\left(Z_{m} - Z_{n}\right)^{q}$, where $m = \textrm{min } (i, j)$ and $n = \textrm{max } (i, j)$. While obtaining a particular $\Delta \Phi_{mn}$, we have accounted for the phase discontinuity $\Phi^{\pm}_{mn}$ (second to last column) in the corresponding crossing: $(i\uparrow / \downarrow, j\uparrow / \downarrow)$.}
\label{Phase_Jump}
\end{table}

In this appendix, we demonstrate that the phase factor acquired by the wavefunction \eqref{StateAlt} upon counterclockwise (clockwise) exchange of the $i$th and the $j$th anyon is indeed $e^{+ i\pi q} \left(e^{- i\pi q}\right)$. For the exchange process, although the anyons start from their respective lattice sites and end at the exchanged lattice sites, they need to move adiabatically and continuously, i.e., we need to change the positions of the $i$th and the $j$th anyons ($Z_{i}$ and $Z_{j}$ respectively) on a continuous path on the plane. This in turn implies that the phase in this process is the monodromy of the Jastrow factor of the continuum wavefunction \eqref{StateContExact}
\begin{equation}\label{NonGauss}
\widetilde{\psi} = \prod_{i<j}(Z_{i}-Z_{j})^{q}.
\end{equation}
Therefore, to determine the exchange phase, we have to determine the change in the phase angles in the individual complex numbers $(Z_{i}-Z_{j})^{q}$ in \eqref{NonGauss}. In particular, we need to compensate for the discontinuous branch cut crossings.

We first consider the case of three anyons and show the intermediate steps of the different exchanges in Fig.\ \ref{fig:phase_jump}. Note that we can move the anyons horizontally without any of the anyons crossing a branch cut, and we can therefore assume that the anyons are initially arranged on a vertical line.

Consider the crossing $(3 \downarrow, 2 \uparrow)$  in the counterclockwise exchange of anyons $\#1$ and $\#3$ keeping anyon $\#2$ to the right of the exchange loop [Fig.\ \ref{fig:phase_jump}(a)]. When anyon $\#3$ is right above $\#2$, the phase of $(Z_{2}-Z_{3})^{q}$ is $\approx 2\pi q$. On the other hand, right after anyon $\#3$ crosses $\#2$, the phase of $(Z_{2}-Z_{3})^{q}$ is $\approx 0$. We need to remember that the vector $(Z_{2}-Z_{3})$ has made a complete $2\pi$ turn. As a result, we have
\begin{equation}\label{IndPhase1}
(Z_{2}-Z_{3})^{q}_\textrm{F} = e^{i\pi q}(Z_{2}-Z_{3})^{q}_\textrm{I},
\end{equation}
where the subscripts `I' and `F' denote initial and final, respectively.

The crossing $(1 \uparrow, 2 \downarrow)$ in the same exchange (a), however, does not lead to a discontinuous jump in the phase angle of $(Z_{1}-Z_{2})^{q}$. Thus, comparing the initial and the final positions of the vector $(Z_{1}-Z_{2})$, we have
\begin{equation}\label{IndPhase1}
(Z_{1}-Z_{2})^{q}_\textrm{F} = e^{-i\pi q}(Z_{1}-Z_{2})^{q}_\textrm{I}.
\end{equation}

More generally, from Table \ref{Phase_Jump}, we observe that for $m<n$ (both positive integers), the crossing $(n \downarrow, m \uparrow)$ leads to a phase discontinuity of $-2\pi q$. Exchanging the arrows, i.e., reversing the directions the anyons are moved, produces a discontinuity $+2\pi q$. For all the instances, where anyon $\#m$ is to the left when crossing anyon $\#n$, the phase angle of $\left(Z_{m} - Z_{n}\right)^{q}$ changes continuously.

The numbering of the anyons in Fig.\ \ref{fig:phase_jump} obeys the rules delineated in Fig.\ \ref{LatticeOrdering} and Sec.\ \ref{sec:model}. We then consider both clockwise and counterclockwise exchanges between anyons $\#1$ and $\#3$. We also consider the distinct situations, where anyon $\#2$ remains either to the left or to the right of the exchange loop. Taking into account all the phase changes listed in Table \ref{Phase_Jump}, we obtain the following relations between the final and the initial wavefunctions:
\begin{subequations} \label{FinExcPhase}
\begin{multline}
\textrm{\textbf{Counterclockwise exchange:}} \quad \widetilde{\psi}_\textrm{F} =  \\
= e^{\mp i\pi q} (Z_{1}-Z_{2})^{q}_\textrm{I} e^{+i\pi q} (Z_{1}-Z_{3})^{q}_\textrm{I} e^{\pm i\pi q} (Z_{2}-Z_{3})^{q}_\textrm{I} = \\
= e^{+i\pi q} \widetilde{\psi}_\textrm{I},
\end{multline}
\begin{multline}
\textrm{\textbf{Clockwise exchange:}} \quad \widetilde{\psi}_\textrm{F} =  \\
= e^{\pm i\pi q} (Z_{1}-Z_{2})^{q}_\textrm{I} e^{-i\pi q} (Z_{1}-Z_{3})^{q}_\textrm{I} e^{\mp i\pi q} (Z_{2}-Z_{3})^{q}_\textrm{I} = \\
= e^{-i\pi q} \widetilde{\psi}_\textrm{I}.
\end{multline}
\end{subequations}
From Eq.\ \eqref{FinExcPhase}, we also observe that the exchange phase is not affected by the position of the other anyon (not involved in the exchange) with respect to the exchange loop. We get the correct phase as long as the other anyon is not enclosed in the exchange loop.

Let us now consider an arbitrary number of anyons. We exchange the $i$th and the $j$th anyons, where $i$ and $j$ are two arbitrary positive integers and $i < j$. The $i$th and $j$th anyons cross the $k$th anyon, with $i<k<j$, either on the left or on the right. As long as no other anyon is enclosed in the counterclockwise (clockwise) exchange loop, similar to Table \ref{Phase_Jump}, we have
\begin{subequations}
\begin{align}
\Delta \Phi_{ij} &= \pi q \; \left(-\pi q\right), \\
\Delta \Phi_{ik} &= -\Delta \Phi_{kj}, \qquad i<k<j. \label{Ind_X2}
\end{align}
\end{subequations}
Therefore, at the end of the counterclockwise (clockwise) exchange, we are left with an exchange phase of $e^{+i\pi q} \left(e^{-i\pi q} \right)$. This justifies our claim  in Sec.\ \ref{sec:model} below Eq.\ \eqref{StateAlt} about the additional phase factor due to the exchange of two anyons.

\begin{acknowledgments}
The authors would like to thank J.\ Ignacio Cirac and Benedikt Herwerth for discussions and Douglas Lundholm for comments on the manuscript. J.W.\ thanks NSF DMR 1306897 and NSF DMR 1056536 for partial support.
\end{acknowledgments}

\end{document}